\documentclass{article}
\usepackage{spconf,epsfig}
\usepackage{amsmath,amsfonts}
\usepackage{algorithmic}
\usepackage{algorithm}
\usepackage{array}
\usepackage[caption=false,font=normalsize,labelfont=sf,textfont=sf]{subfig}
\usepackage{xcolor}
\usepackage{textcomp}
\usepackage{stfloats}
\usepackage{url}
\usepackage{verbatim}
\usepackage{graphicx}
\usepackage{multirow}
\usepackage{setspace}

\DeclareMathOperator*{\argmax}{arg\,max}

\title{Graph-based Active Learning for Surface Water and Sediment Detection in Multispectral Images}
\name{Bohan Chen$^1$\sthanks{Supported by the UC-National Lab In-Residence Graduate Fellowship Grant L21GF3606}, Kevin Miller$^2$\sthanks{Supported by the Department of Defense (DOD) National Defense Science and Engineering Graduate (NDSEG) Fellowship}, Andrea L. Bertozzi$^1$\sthanks{ALB is supported by the 
National Geospatial-Intelligence Agency under Award No.
HM0476-21-1-0003. Any opinions, findings and conclusions or recommendations expressed in this material are those of the authors and do not necessarily reflect the views of the National Geospatial-Intelligence Agency. Approved for public release, NGA-U-2023-01028.},
Jon Schwenk$^3$\sthanks{Supported by the Laboratory Directed Research and Development program of Los Alamos National Laboratory under project numbers 20170668PRD1 and 20210213ER}}
\address{$^1$ University of California, Los Angeles. Math Dept. 520 Portola Plaza, Los Angeles, CA 90095, USA.\\
$^2$ University of Texas, Austin. Oden Institute for Computational Engineering and Sciences.\\
201 E 24th St, Austin, TX 78712, USA.\\
$^3$ Los Alamos National Laboratory. Earth \& Env. Sci. Division. Los Alamos, NM 87545, USA.}
    
\begin{document}
%
\maketitle
\begin{abstract}
We develop a graph active learning pipeline (GAP) to detect surface water and in-river sediment pixels in satellite images. The active learning approach is applied within the training process to optimally select specific pixels to generate a hand-labeled training set. Our method obtains higher accuracy with far fewer training pixels than both standard and deep learning models. According to our experiments, our GAP trained on a set of 3270 pixels reaches a better accuracy than the neural network method trained on 2.1 million pixels.
\end{abstract}

\section{Introduction}\label{sec:intro}
Surface water dynamics are critical to climate, flood monitoring and mitigation, freshwater resource management, water quality analyses, and geomorphology \cite{bates2013observing, cooley2021human, chen2021estimating}. 
 The importance of automated surface water detection in remote sensing is highlighted by the publication of global datasets or pre-trained surface water models, such as the Global Surface Water dataset \cite{pekel2016high}, Surface Water Extent product \cite{jones2019improved}, and DeepWaterMap \cite{isikdogan2019seeing}, in combination with different classification methods including support vector machine (SVM), random forest (RF), and convolutional neural networks (CNN). 
 

Beyond just surface water, some river studies require mapping of in-channel sediment to identify their so-called ``bankfull'' state \cite{bjerklie2007estimating}, which we define as the union of water and active (unvegetated) in-channel sediment bars \cite{schwenk2017high}. We build and employ our model on the small but high-quality, manually-labeled RiverPIXELS dataset \cite{schwenk2022riverpixels} consisting of labeled water and in-river, unvegetated sediment from multispectral Landsat images.

Our main contribution is a graph-based active learning pipeline (GAP) to identify water and sediment pixels in multispectral images. The inclusion of the sediment class has largely been omitted in previous efforts or treated as an independent problem (e.g., \cite{monegaglia2018automated}). 
Our model is based on the RiverPIXELS dataset to classify water and in-channel sediment pixels in multispectral images. The first challenge of this problem is the paucity of training data. The RiverPIXELS dataset does not have enough images to train an accurate CNN model to classify pixels. We address this insufficiency by graph learning, which is good at classification tasks with a low label rate \cite{miller2022graph}. The second challenge is designing an efficient graph learning approach. The RiverPIXELS dataset contains millions of labeled pixels. Using all available pixels would result in extreme computational demand with long runtimes. We implement an active learning approach to condense the training set by selecting representative pixels accounting for approximately 0.1\% of the total number of labeled pixels. 
Compared to methods in other products, our GAP achieves the highest accuracy with minimal training data (i.e., the number of labeled pixels). All our codes are available on GitHub\footnote{\url{https://github.com/wispcarey/SurfWater-Graph-Active-Learning-GAP-}}.

\section{Graph Active Learning Pipeline}
\label{sec:GAP}

\begin{figure*}[!t]
\centering
\includegraphics[width=7in]{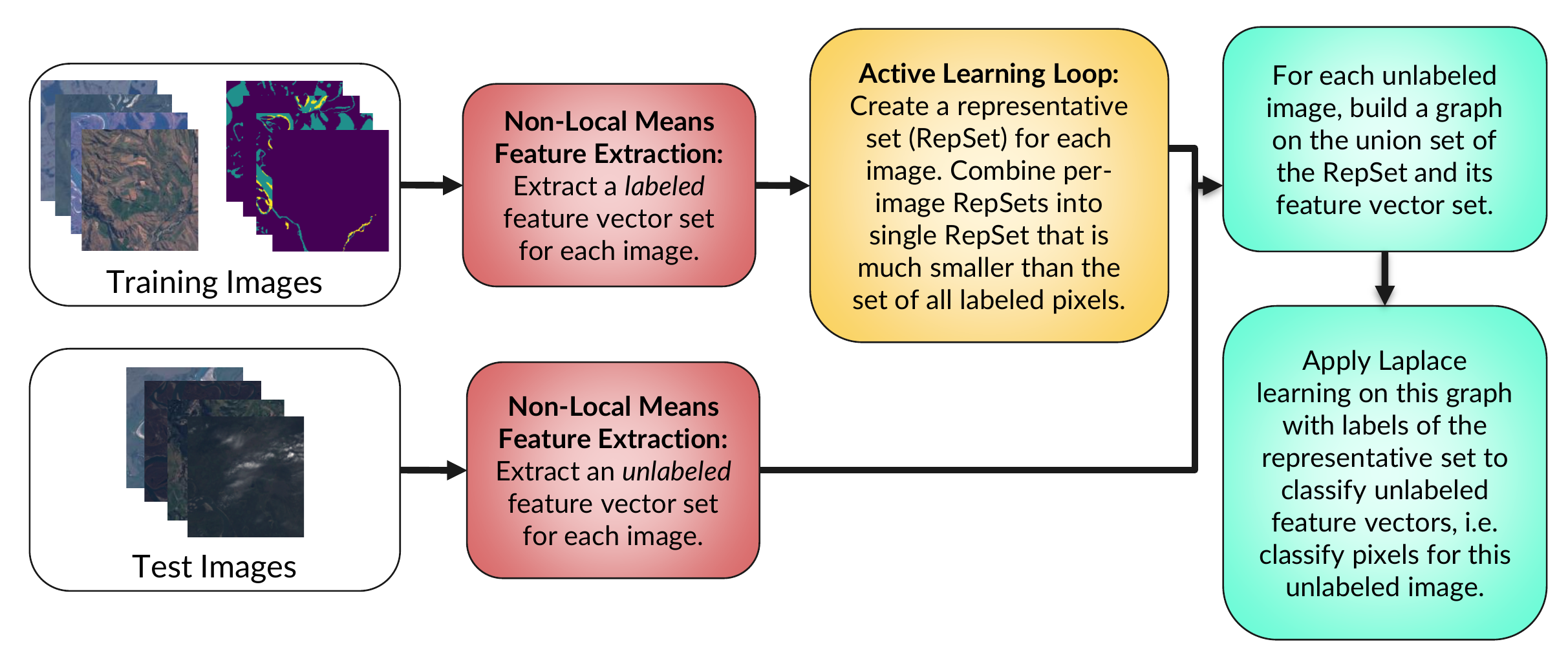}
\vspace{-0.8cm}
\caption{\small The flowchart of our GAP: 1. (Red Boxes) Apply the non-local means method to extract feature vectors. 2. (Yellow Box) Condense the labeled feature vector set into a smaller representative set (RepSet) through active learning. 3. (Cyan Box) Build a graph based on the union of the RepSet and the unlabeled feature set. Then, apply graph learning approaches to predict labels for unlabeled feature vectors.}
\label{fig:basic_pipeline}
\end{figure*}

The graph active learning pipeline (GAP) is summarized in Figure~\ref{fig:basic_pipeline}). We extract a non-local means feature vector corresponding to a neighborhood of each pixel in a given image, then construct a similarity graph $G=(X,W)$ based on the feature vectors $X = \{x_1,x_2,\ldots,x_N\}$, where $W$ is the edge weight matrix generated according to KNN angular similarity. With the ground-truth labels on a subset $X_L\subset X$, we apply graph Laplace learning \cite{zhu2003semi} to predict the classification of the unlabeled nodes. Active learning approaches are applied to select the labeled set $X_L$ based on the Model-Change (MC) acquisition function $\mathcal{A}: X-X_L\rightarrow \mathbb{R}$ \cite{miller2020efficient, miller2021model}. A similar method for an image segmentation task is introduced in \cite{Chen2023Batch}.

The first step to classifying pixels is to generate a feature vector for each pixel. For pixel $i$, consider a $(2k+1)\times (2k+1)$ neighborhood patch centered at pixel $i$ (for boundary pixels, use the reflection padding). Inspired by the non-local means method \cite{buades2005non}, we use the flattened Gaussian weighted neighborhood patch as the non-local means feature vector \cite{Chen2023Batch} of pixel $i$. Then we extract the non-local means feature vector for each image in the training image set $\mathcal{I}$ and select a subset of labeled feature vectors as our training set.


Selecting a reasonable $X_L$ is tricky. Each image in RiverPIXELS is of the size $256\times 256$, and the whole dataset includes millions of labeled pixels. It is thus inefficient to build a graph using all of the labeled pixels. Instead, we apply an active learning approach to select $X_L=R$, where $R$ is a subset of representative labeled feature vectors, called the representative set (RepSet). We find a representative set $R_{j}$ for each image $I_{j}\in\mathcal{I}$ and combine them together to obtain $R = \cup_{i=1}^{l_n} R_{i}$. With the following four steps (Algorithm~\ref{alg:create_rep}), we determine the representative set for a certain image $I_{i} \in \mathcal{I}$:


\begin{algorithm}[t!]
\caption{Create the Representative Set}\label{alg:create_rep}
\small
\begin{algorithmic}
\STATE {\textsc{Input:}} Image set $\mathcal{I} = \{I_1,I_2,\ldots,I_l\}$. 
\STATE {\textsc{Output:}} Representative set $\mathcal{R}$ of the training image set $\mathcal{I}$.
\STATE $\mathcal{R}\gets \emptyset$; $i\gets 1$ 
\STATE {\textsc{For}} $i=1$ to $l$, {\textsc{Do}}:
\STATE \hspace{0.5cm}\textbf{Preprocess} image $I_i$ to construct the feature vector of each pixel
\STATE \hspace{0.5cm}\textbf{Initialize} the representative set $R_i^0$ for image $I_i$ 
\STATE \hspace{0.5cm}\textbf{Active learning loop} until \textbf{Termination} to obtain final $R_i$
\STATE \hspace{0.5cm}$\mathcal{R} \gets \mathcal{R}\cup R_i$
\end{algorithmic}
\end{algorithm}

\noindent\textbf{1. Preprocess:} Apply the non-local means method to get the feature set $X_i = \{x^i_j\}$ of image $I_i$. 
    
\noindent\textbf{2. Initialization:} Initialize the representative set $R_i^0$ by randomly selecting the same number of feature vectors of each class from the feature set $X_i$. This initialization requires all ground-truth labels of feature vectors in $X_i$ and gives a class-balanced initial $R_i^0$.
                
\noindent\textbf{3. Active learning loop:} Using active learning approaches, add feature vectors from $X_i$ and corresponding labels one by one to the representative set (Algorithm~\ref{alg:Active learning loop}).

\begin{algorithm}[t!]
\small
\caption{Active learning loop}\label{alg:Active learning loop}
\begin{algorithmic}
\STATE {\textsc{Input:}} The initialized representative set $R^0$ and feature vector set $X$.
\STATE {\textsc{Output:}} The representative set $R$ of the feature vector set $X$.
\STATE Build a graph $G$ on $R^0\cup X$; $t \gets 0$
\STATE {\textsc{While}} \textbf{terminal condition} is \textsc{not} triggered:
\STATE \hspace{0.5cm} Apply Laplace learning on $G$ with labels of $R^t$
\STATE \hspace{0.5cm} Get the predicted labels of $X$
\STATE \hspace{0.5cm} Calculate the acquisition function $\mathcal{A}(x_k),\ \forall x_k\in X\setminus R^0$
\STATE \hspace{0.5cm} $k^\ast\gets\argmax_{x_k\in X\setminus R^0}\mathcal{A}(x_k)$
\STATE \hspace{0.5cm} $R^{t+1}\gets R^t\cup\{x_{k^\ast}\}$; $t \gets t + 1$
\STATE $R\gets R^t$
\end{algorithmic}
\end{algorithm}

\noindent\textbf{4. Termination:} Stop the active learning loop when a certain termination condition is satisfied. The accuracy-based terminal condition is applied. In iteration $t$ of the \textbf{active learning loop} step for image $I_i$, we apply Laplace learning based on labels of the current representative set $R_i^{t}$ to predict other nodes $X_i-R_i^{t}$. Denote the prediction accuracy on $X_i-R_i^{t}$ by $a_t$. Given a positive integer $K_\text{max}$ and a positive real number $\epsilon>0$, the active learning loop terminates if $|a_t-a_{t-1}| <\epsilon \text{ or } t > K_\text{max}$.

In summary, we extract non-local means feature vectors for the training images, then apply the active learning method to condense them into a small RepSet $R$. For each test image $\tilde{I}_i$ (considered unlabeled), we combine its extracted feature vector set $\tilde{X}_i$ with $R$ to build a graph. Finally, we apply graph Laplace learning to predict labels on $\tilde{X}_i$. This node classification gives the segmentation of image $\tilde{I}_i$.


\section{Experiments and Results}\label{sec: experiments}
We choose five rivers from the RiverPIXELS dataset: the Kolyma, Yana, Waitaki, Colville, and Ucayali Rivers. Our pipeline is trained on a small subset of pixels chosen from the first four rivers, while the performance of different methods is tested on the remaining pixels of all five rivers. There are $42$ images belonging to the first four river regions, while the Ucayali River includes $54$ images. We randomly sample $75\%$ of the labeled data ($32$ images) for each region as set $\mathcal{I}$ from which we develop the smaller training set (RepSet) $R$ and use the remaining $25\%$ ($10$ images) as the test set $\tilde{\mathcal{I}}$. In addition, we employ an extra test set $\tilde{\mathcal{I}}_\text{ex}$ formed by $54$ images of the Ucayali river. Of note is that the Ucayali river is in the tropical Amazonian region, while the other four are within temperate or arctic regions, creating an additional challenge for the method. 

Model performances are evaluated using metrics of the \textbf{Boundary Accuracy} and \textbf{Overall Accuracy}. We define a pixel's boundary distance $D_B$ as the minimal distance to a pixel with a different ground-truth label. The \textbf{Boundary Accuracy} $BA(d)$ is defined as the accuracy on pixels with a boundary distance $D_B \leq d$. The \textbf{Overall Accuracy} (OA) is the accuracy on all pixels. BA is more indicative of the model performance than OA since a naive classifier that classifies all pixels into the land will have an OA of around 80\%.

We compare the classification performance of our GAP to DeepWaterMap (DWM) \cite{isikdogan2019seeing}, support vector machine (SVM) \cite{cortes1995support} and random forest (RF) \cite{ho1995random} models. For our GAP, we extract $7\times 7$ non-local means features of pixels in the training image set. The original extracted feature set $X\subset \mathbb{R}^{294}$ has over 2 million feature vectors from the training set consisting of 32 labeled images. DWM only provides the classification of water and land pixels, while RiverPIXELS images include water, bare sediment, and land. We provide two approaches to compare the performance between other methods and DWM. The first approach is to \textit{retrain DeepWaterMap} (DWM\_R). We train a new neural network with the same structure of DWM on our training set with 32 labeled images and labels of water, sediment, and land. The second approach is to \textit{modify labels}. Inspection of the training set of the original DeepWaterMap (DWM\_O) shows that nearly all sediment pixels are labeled as land. We modify the labels of our training set and the ground-truth labels of our test set by changing sediment labels to land labels. Each method uses a different amount of training data chosen to optimize the performance of each method. For SVM and RF, we use a training set named T-NLM consisting of 42.6K labeled non-local means feature vectors randomly sampled and balanced in each class. The retrained DWM (DWM\_R) is trained on $\mathcal{I}$ while the original DWM (DWM\_O) is trained on around 100K labeled images.

\begin{figure*}[!t]
\centering

\subfloat[RGB]{\includegraphics[width=1.1in]{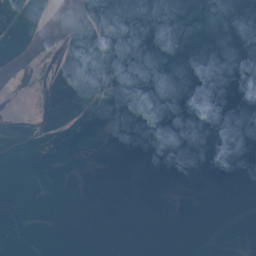}%
}
\hfill
\subfloat[Ground-truth]{\includegraphics[width=1.1in]{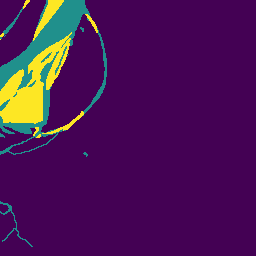}%
}
\hfill
\subfloat[B-GAP (ours)]{\includegraphics[width=1.1in]{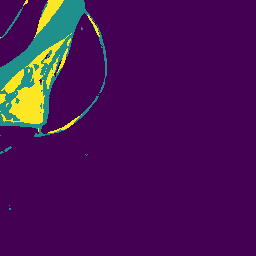}%
}
\hfill
\subfloat[DWM]{\includegraphics[width=1.1in]{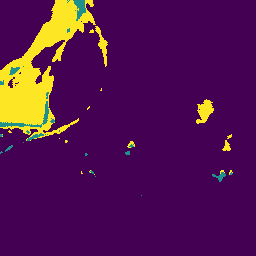}%
}
\hfill
\subfloat[SVM]{\includegraphics[width=1.1in]{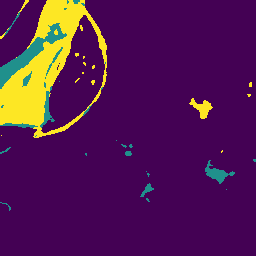}%
}
\hfill
\subfloat[RF]{\includegraphics[width=1.1in]{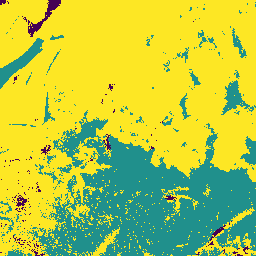}%
}

\caption{\small Results for a Patch of Ucayali river. Original Patch name: \textit{Ucayali\_River\_1 2018-09-11 006 066 L8 316 landsat}. This image has some clouds on the top. In panel (b)-(f), purple, blue, and yellow are land, water, and sediment respectively.}
\label{fig: Ucayali_image}
\end{figure*}

Table~\ref{tab:compare_10} shows our experiment results, including the training and test set information and the BA(3), BA(10), and OA accuracy metrics under both the retrained DWM method and the modified label (Sed$\rightarrow$Land). Note that our GAP method is trained on the smallest training set yet has the highest boundary accuracies, BA(3) and BA(10), as well as the highest overall accuracy on both the test set $\tilde{\mathcal{I}}$ and the extra test set $\tilde{\mathcal{I}}_\text{ex}$. Note that our GAP method still performs the best across different regions (extra test set). Figure~\ref{fig: Ucayali_image} displays results for a sampled image from the Ucayali river that includes cloud cover. 

\begin{table*}[!t]
\begin{center}
\resizebox{\textwidth}{!}{
\begin{tabular}{|c||c|c||c|c||c|c|c||c|c|c||}
\hline
\multirow{2}{*}{\textit{\textbf{Method}}}&\multicolumn{2}{c||}{\textit{\textbf{Training Set}}} &\multicolumn{2}{c||}{\textit{\textbf{Test Set}}} &\multicolumn{3}{c||}{\textit{\textbf{Retrain DWM}}(\%)} &\multicolumn{3}{c||}{\textit{\textbf{Sed$\rightarrow $Land}}(\%)}\\\cline{2-11}
&\textit{Dataset} &\textit{Size} &\textit{Dataset} &\textit{\# of Images} &\textit{BA(3)} &\textit{BA(10)} &\textit{OA} &\textit{BA(3)} &\textit{BA(10)} &\textit{OA} \\\hline\hline

GAP (ours)&RepSet&3.27K Pixels&\multirow{5}{*}{$\tilde{\mathcal{I}}$}&\multirow{5}{*}{$10$}&\textbf{81.90} &\textbf{90.41} &\textbf{95.50}& \textbf{83.75} &\textbf{91.66} &\textbf{96.30}\\\cline{1-3}\cline{6-11}
SVM&T-NLM&42.6K Pixels&&&79.28 &88.05 &93.55& 82.77 &91.08 &95.41\\\cline{1-3}\cline{6-11}
RF&T-NLM&42.6K Pixels&&&77.40 &86.84 &93.31& 81.50 &89.63 &95.07\\\cline{1-3}\cline{6-11}
DWM\_R&$\mathcal{I}$&32 Images&&&72.56 &84.56 &92.86& N$\backslash$A& N$\backslash$A& N$\backslash$A\\\cline{1-3}\cline{6-11}
DWM\_O&$\mathcal{I}_\text{DWM}$&100K Images&&&N$\backslash$A& N$\backslash$A& N$\backslash$A& 78.02 &88.81 &95.11\\\hline\hline

GAP (ours)&RepSet&3.27K Pixels&\multirow{5}{*}{$\tilde{\mathcal{I}}_\text{ex}$}&\multirow{5}{*}{$54$}&\textbf{83.07} &\textbf{92.93} &\textbf{97.48}& \textbf{85.08} &\textbf{94.17} &\textbf{98.21}\\\cline{1-3}\cline{6-11}
SVM&T-NLM&42.6K Pixels&&&79.08 &90.29 &96.26& 77.49 &90.96 &97.23\\\cline{1-3}\cline{6-11}
RF&T-NLM&42.6K Pixels&&&26.41 &26.79 &12.52& 76.01 &89.09 &96.21\\\cline{1-3}\cline{6-11}
DWM\_R&$\mathcal{I}$&32 Images&&&69.28 &83.43 &94.39& N$\backslash$A& N$\backslash$A& N$\backslash$A\\\cline{1-3}\cline{6-11}
DWM\_O&$\mathcal{I}_\text{DWM}$&100K Images&&&N$\backslash$A& N$\backslash$A& N$\backslash$A& 79.42 &91.14 &97.29\\\hline
\end{tabular}}
\end{center}
\vspace{-0.5cm}
\caption{\small Comparison of different methods (GAP, SVM, RF, DWM\_R, and DWM\_O), where each is evaluated by boundary accuracies BA(3), BA(10), and the overall accuracy (OA).
The training set size of our GAP, SVM, and RF is the number of pixels used in the training set, while that of DWM\_R and DWM\_O is the number of images in the training set since DWM is a CNN-based method. Each image contains $65536$ pixels. The performance of different methods is evaluated on both the test set $\tilde{\mathcal{I}}$ and the extra test set $\tilde{\mathcal{I}}_\text{ex}$, in a different region from the training set $\mathcal{I}$. Our GAP method has the best performance according to these accuracy values.}
\label{tab:compare_10}
\end{table*}

\section{Conclusion}
We develop a graph active learning pipeline (GAP) to classify pixels into the land, water, and bare sediment classes from the RiverPIXELS dataset. GAP outperforms the classical methods of support vector machine and random forest, and a cutting-edge CNN approach (DeepWaterMap) trained on a dataset hundreds of times larger than RiverPIXELS. With the help of active learning techniques, GAP can be trained on much smaller datasets than other methods. Furthermore, GAP demonstrates superior performance on images of rivers in different environments relative to other approaches. 
Our work demonstrates how graph-based active learning techniques can provide efficient labeling of and greater accuracy in detecting surface water and in-river sediment using significantly fewer training data than classic and deep approaches.

\begingroup
\setstretch{0.5}
\bibliographystyle{IEEEbib}
{\small
\bibliography{refs}}
\endgroup

\end{document}